\documentclass[3p,times,twocolumn]{elsarticle}

\usepackage{ecrc}

\volume{00}

\firstpage{1}

\journalname{Nuclear Physics B Proceedings Supplement}

\runauth{}

\jid{Strongly Coupled}

\jnltitlelogo{Nuclear Physics B Proceedings Supplement}

\usepackage{amssymb}

\usepackage[figuresright]{rotating}

\begin{document}

\begin{frontmatter}





\title{Strongly Coupled Cosmologies}

\author[a1,a2]{S.A.~Bonometto}
\author[a1,a2]{M.~Mezzetti}
\author[a3]{I.~Musco}
\author[a4]{R.~Mainini}
\author[a5]{A.V.~Macci\`o}

\address[a1]{Trieste University, Physics Department, Astronomy Unit,
  Via Tiepolo 13, I34143 Trieste, Italy} \address[a2]{I.N.A.F.,
  Trieste Observatory, Via Tiepolo 13, I34143 Trieste, Italy}
\address[a3]{Laboratoire Univers et Théories, UMR 8102 CNRS, Observatoire 
de Paris, Universit\'e Paris Diderot, 5 Place J.~Janssen, 92190 Meudon, France} 
\address[a4]{Milano--Bicocca University, Physics Department
  G.~Occhialini, Piazza della Scienza 3, I20126 Milano, Italy}
\address[a5]{ Max-Planck-Institut f\"ur Astronomie, K\"onigstuhl 17,
  D69117 Heidelberg, Germany}

\begin{abstract}

Models including an energy transfer from CDM to DE are widely
considered in the literature, namely to allow DE a significant
high--$z$ density. Strongly Coupled cosmologies assume a much larger
coupling between DE and CDM, together with the presence of an
uncoupled warm DM component, as the role of CDM is mostly restricted
to radiative eras. This allows us to preserve small scale fluctuations
even if the warm particle, possibly a sterile $\nu$, is quite light,
$\cal O$$(100$eV). Linear theory and numerical simulations show that
these cosmologies agree with $\Lambda$CDM on super–galactic scales;
e.g., CMB spectra are substantially identical.  Simultaneously,
simulations show that they significantly ease problems related to the
properties of MW satellites and cores in dwarfs. SC cosmologies also
open new perspectives on early black hole formation, and possibly lead
towards unificating DE and inflationary scalar fields.

\end{abstract}





\end{frontmatter}

\section{Introduction}
We discuss a new family of models, starting from their features in the
radiative eras. In such epochs, they are chacterized by two extra
components, in top of the {\it usual} radiative ones: a scalar field
$\Phi$ and a {\it peculiar} CDM component, with energy densities and
pressures $\rho_\Phi$ \& $\rho_c$ and $p_\phi$ \& $p_c$,
respectively. As we assume $\rho_\Phi \simeq p_\Phi \simeq \dot
\Phi^2/2 a^2$ and being $p_c \simeq 0$, it should be $\rho_\Phi
\propto a^{-6}$, $\rho_c \propto a^{-3}$. It is not so, because the
models assume an energy flow from CDM to the field, due to a
Yukawa--like interaction Lagrangian
\begin{equation}
{\cal L}_I = -\mu f(\Phi/m) \bar \psi \psi~,
\label{interact}
\end{equation}
$\psi$ being the CDM spinor field. If
\begin{equation}
f = \exp(-\Phi/m)
\label{fff}
\end{equation}
with $m=m_p/b$, there exists a solution with
\begin{equation}
\rho_c \propto f(\Phi/m) a^{-3}
\label{rhoscal}
\end{equation}
\begin{equation}
\Phi = m \ln(\tau/\tau_r)
\label{ansatz}
\end{equation}
being an attractor for the system made by the equations of motions of
$\Phi$ and $\psi$ ($m_p$: the Planck mass, $\tau_r$: a generic
reference conformal time).  Eqs.~(\ref{rhoscal}) and (\ref{ansatz})
imply that $\rho_c \propto \rho_\Phi \propto a^{-4}$, so that CDM and
the field dilute at the same rate of the radiative components.  On the
attractor, the constant early state parameters $\Omega_c $ and
$\Omega_\Phi$ (CDM and $\Phi$, respectively) shall then read
\begin{equation}
\Omega_c = 1/(2\beta^2)~,~~~ \Omega_\Phi = 1/(4 \beta^2)
\label{omegas}
\end{equation}
\begin{figure}
\begin{center}
\vskip -.5truecm
\includegraphics[height=7.5cm,angle=0]{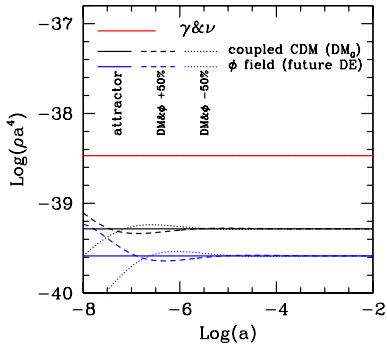}
\end{center}
\vskip -2.3truecm
\caption{Starting from generic initial conditions, the attractor is
  soon recovered: both $\rho_c$ and $\Phi$ were initially ``wrong''
  or, equivalently, the rate of expansion was ``wrong''. Here
  $\beta=2.8$, so that the overall density of CDM and $\Phi$ is
  smaller than half extra $\nu$ species. }
\label{fig1}
\vskip -.3truecm
\end{figure}
with 
\begin{equation}
\beta^2  = (3/16\pi) b^2~.
\label{beta}
\end{equation}
What happens is that the flow of energy from CDM to the field fastens
(slows down) the dilution of CDM ($\Phi$); accordingly, the scale
factor exponents change: for the former component from -3 to -4, for
the latter one from -6 to -4~.  The notable point is that this
behavior occurs along an attractor: if starting from generic initial
conditions, with $\Omega_\Phi$ and/or $\Omega_c$ different from
eq.~(\ref{omegas}), the e.o.m. rebuild the conditions (\ref{omegas}),
also suitably synchronizing the rates of energy transfer and cosmic
expansion.  For a detailed proof, see Paper A \cite{papera}, wherefrom
Figure \ref{fig1} is taken.

These models are then characterized by three phases: (B) before (the
radiative eras); (D) during (them); (A) after (matter--radiation
equality).

The stages (D) and (A) were treated both in \cite{papera} and in a
further Paper B \cite{paperb}, focused on fluctuation evolution,
finding that, from the above attractor, the models naturally evolve
towards a picture consistent with today's Universe. Besides of baryons
this requires a WDM (Warm Dark Matter) component (see, e.g.,
\cite{bono1}).

More in detail: In the present epoch DM is essentially warm, $\Phi$
has turned into quintessential DE (Dark Energy), while CDM, keeping an
almost negligible density, seems to play just an ancillary role.
\begin{figure}
\begin{center}
\vskip -.5truecm
\includegraphics[height=7.5cm,angle=0]{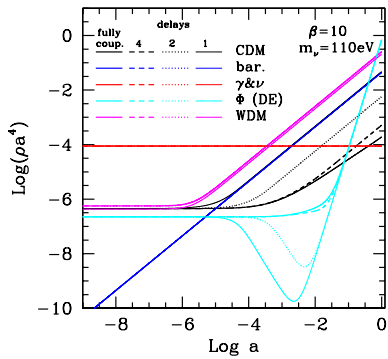}
\end{center}
\vskip -2.3truecm
\caption{Background evolution. WDM derelativisation breaks the former
  conformal invariance, allowed by CDM--$\Phi$ coupling. Different
  curves yield models where coupling either persists down to $z=0$ or
  fades earlier (see text). The $w$ (field state parameter) shift,
  from +1 to -1 is tuned to account for a present DE density
  parameter~$\Omega_d=~0.7$.  }
\label{fig2}
\vskip -.3truecm
\end{figure}

In Figure \ref{fig2} we show the background evolution in some
spatially flat models with $\Omega_d=0.7$, $\Omega_b=0.045$,
$h=0.685$, in agreement with Planck results. Model dynamics is somehow
reminiscent of the coupled DE option, studied by many authors
\cite{cDE}. Here, however, coupling plays its key role through
radiative eras. Switching it off (or letting $m \to \infty$) after WDM
has derelativized, could even ease the fit with data.  Besides of a
coupling persistent down to $z=0$, we therefore consider the cases of
$\beta$ fading exponentially at $z = z_{der} \times 10^{-D}$ (here we
assume that WDM is a sterile $\nu$ with a former thermal distribution
and the redshift $z_{der}$ is when $m_\nu = T_\nu$; the exponent $D$
is dubbed {\it delay}).

\begin{figure}
\begin{center}
\vskip -.5truecm
\includegraphics[height=7.5cm,angle=0]{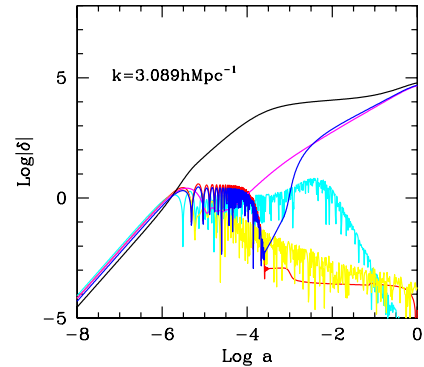}
\end{center}
\vskip -2.1truecm
\caption{Linear fluctuation evolution in a model with $D=2$. The wave
  number considered corresponds to a scale of $8 \, h^{-1}$Mpc. Colors
  as in Figure 2. The scale in ordinate is arbitrary (but see text).
}
\label{fig3}
\vskip -.3truecm
\end{figure}

\section{Fluctuation evolution}
In Paper B it is widely discussed how the public program CMBFAST (or,
similarly, CAMB) is to be modified to follow fluctuation evolution in
these models. Changes involve also out--of--horizon initial
conditions.

A notable feature of these models is that there is no cut in the
transfer function of WDM. The point is that, in the non--relativistic
regime, the presence of a CDM--$\Phi$ coupling yields an increase of
the effective self--gravity of CDM, as though
\begin{equation}
G \to G^* = G(1+4\beta^2/3)
\label{g*}
\end{equation}
(see \cite{maccioamen}). The interaction of CDM with other components
is set by $G$, the $G$--shift concerning just CDM--CDM gravity. When
fluctuations reach the horizon, the CDM density parameter is $\cal
O$$(\beta^{-2})$. However, as its self interaction is boosted by a
factor $\cal O$$( \beta^{2})$, it evolves as though $\Omega_c \sim
2/3$, indipendently of $\beta$.

The other, velocity dependent, changes in CDM dynamics, not discussed
here, do not modify the fact that the growth of CDM fluctuations is
never dominated by the baryon--radiation plasma. Accordingly, while
baryons and radiation yield sound waves, and collisionless components
suffer free streaming, the CDM fluctuations $\delta_c$ suffer no {\it
  Meszaros}'s effect and steadily grow. Later on, when WDM
derelativizes and/or baryons decouple from radiation, $\delta_c$ has
grown so large, to cause the re--generation of fluctuations in the
other components, and this is an essential feature to meet specific
data $\Lambda$CDM models do not fit.

\begin{figure}
\begin{center}
\vskip -.5truecm
\includegraphics[height=7.5cm,angle=0]{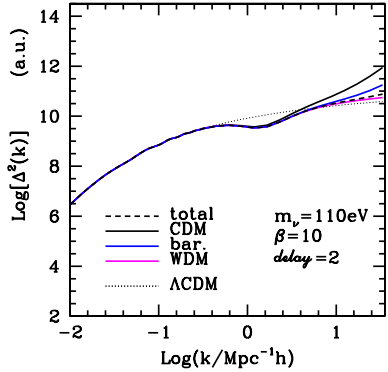}
\end{center}
\vskip -2.1truecm
\caption{Linear spectral function in a model with $D=2$. The scale in
  ordinate is arbitrary. The dashed thin line is the spectral function
  for a $\Lambda$CDM model with the same normalization. }
\label{fig4}
\vskip -.3truecm
\end{figure}

In Figure~\ref{fig3} we give an example of the linear evolution of
density fluctuations down to $z=0$. The spectrum
\begin{equation}
\Delta^2(k) = {1 \over 2 \pi^2} k^3 P(k)
~~~~~ {\rm with} ~~ P(k) = \langle |\delta|^2 \rangle
\label{spectra}
\end{equation}
is then shon in Figure \ref{fig4} at $z=0$, as obtainable from the linear
code, and with arbitrary normalization. The model considered has
$D=2$; CDM--CDM interactions are therefore ruled by the normal $G$
since a redshift $z \sim 10^3.$
Let us outline, first of all, that different components exhibit
different spectra, even at $z=0$, with CDM showing the most peculiar
behavior. While, in the relativistic regime, outside the horizon, CDM
fluctuations behave similarly to other components, later on, between
the entry in the horizon and $z \simeq 10^{-2}z_{der} $, they undergo
an authonomous fast growth. The entry in the horizon and, therefore,
the duration of this growth depends on mass scale. Figure \ref{fig4}
shows a CDM spectral function exceeding the total function by one
o.o.m. already at $k \simeq 40\, h$Mpc$^{-1}$, corresponding to a mass
scale $M \simeq 9 \times 10^{8} M_\odot h^{-1}.$

The distinction between baryon and CDM spectra, in Figure \ref{fig4},
however, is artificious, being due to the algorithm used. After
CDM--$\Phi$ decoupling baryons and CDM fulfill the same linear
equations, so that possible non--linearities depend on the amplitude
of the weighted sum $\Omega_c\delta_c + \Omega_b \delta_b$. Of course,
there are mass scales where CDM is already non--linear at
$10^{-D}z_{der} $, for any $D > 0$; for the model in Fig.~\ref{fig4}
this occurs for $M < \sim 10^6 M_\odot h^{-1}$. If CDM
non--linearities arise, the code is scarsely predictive, even though
we may conjecture its results to be reasonnably well approximated, for
the non--CDM components, until $\delta_c <\sim 10$ (approximately a
{\it spherical} fluctuation turnover). This does not necessarily mean
that the model is unphysical.

\section{Dwarf galaxy cores and MW satellites}
\begin{figure}
\begin{center}
\vskip -.5truecm
\includegraphics[height=7.5cm,angle=0]{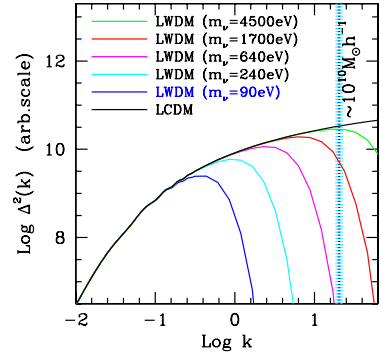}
\end{center}
\vskip -2.1truecm
\caption{Linear spectral function in $\Lambda$WDM models with various
  neutrino masses, as indicated in the frame. }
\label{fig5}
\vskip -.3truecm
\end{figure}

The key issue, however, is the similarity between this model and
$\Lambda$CDM. Plotting $\Delta^2(k)$ aims to stress model
discrepancies, less evident in transfer functions or integral
quantities (e.g.~$\sigma_R$). $\Lambda$CDM is a benchmark for any
model improvement, its main discrepancies from data being: (i) the
number of MW satellites \cite{MWsat}; (ii) (dwarf) galaxy cores
\cite{cores}. To overcome such difficulties, the option of replacing
cold with warm DM has been explored. According to \cite{maccio2012},
however, dwarf galaxies in $\Lambda$WDM N--body simulations exhibit a
core radius
\begin{equation}
R_{core} \sim 1\, h^{-1}{\rm kpc}\, (100\, {\rm eV}/m_\nu)^{1.8}~.
\label{core}
\end{equation}
Observations require $R_{core} \sim 0.5$--1$\, h^{-1}$kpc;
$\Lambda$WDM cosmologies, therefore, are no solution unless $m_\nu
\sim 100\, $eV. In Figure \ref{fig5} we then show linear spectral
functions for a number of $m_\nu$ values, showing that
\begin{figure}
\begin{center}
\vskip -1.3truecm
\includegraphics[height=7.5cm,angle=0]{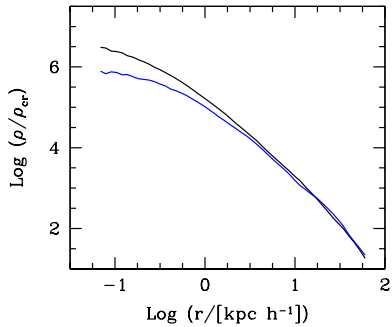}
\end{center}
\vskip -2.4truecm
\caption{Profile of a bound system of mass $\sim 6 \times
  10^{10}M_\odot h^{-1}$ in N--body simulation of a SC cosmology (in
  blue; here $m_\nu=90\, $eV) compared with the same system in a
  $\Lambda$CDM model (in black). }
\label{profile}
\vskip -.5truecm
\end{figure}
a WDM cosmology with $m_\nu \sim 100\, $eV yields no systems even with
mass $\sim 10^{13} M_\odot h^{-1}$. In the literature, it is often
assumed that cosmological data suggest a sterile $\nu$ with mass
$m_\nu \sim 2$--3$\, $keV. According to Fig.~\ref{fig5}, such $\nu$ is
barely sufficient to produce structures over galactic scales, but
yields no improvement on the above problems (i),(ii). Within SC
cosmologies, on the contrary, {\bf we predict a sterile neutrino with mass
  $\cal O$$(100\, $eV).} 

N--body simulations for these cosmologies are in progress and suggest
that, besides of yielding a fair core profile (see preliminary results
in Figure \ref{profile}) such models approximately half the prediction
of MW satellites.

\section{Discussion}
Besides of easing these problems, SC cosmologies open two significant
perspectives. Let us then briefly comment on the impact of the fast
growth of CDM fluctuations on small scale structures, as well as on
the stages (B), i.e. before the onset of radiative expansion.

The increase of $\Delta^2(k)$ towards large $k$'s tells us that, even
in $\Lambda$CDM cosmologies we predict the onset of non--linearity at
scales $< R_{nl}(z)$, at any redshift $z$. This is due to the (slow)
CDM fluctuation growing even in the presence of Meszaros' effect. SC
cosmologies are surely characterized by a much more rapid CDM
fluctuation growth. In any SC model, therefore, CDM fluctuations on
scales $<\sim 10^4 M_\odot h^{-1}$ get non linear before radiative
expansion ends. The rate of a spherical collapse can also be roughly
estimated and, if virial equilibrium is not reached, it could approach
a relativistic regime.

This suggests us two specific comments: (i) during pre--relativistic
non--linear stages, the coherence between CDM and other component
distributions might fade over small scales, causing a low mass
cut--off in the transfer function, however much below the one in
Figure \ref{fig5}; (ii) when approaching a relativistic regime, the
approximated expression (\ref{g*}) looses validity. Let us recall that
a cosmological constant $\Lambda$ was already found to increase the
threshold for cosmological BH formation \cite{MM}. The problem here is
more intricated, however: a $\Phi$ component is already significant at
high $z$; a possible conjecture is that BH formation is also
suppressed, at least until $\beta$--coupling is active. The time and
space dependence of $\Phi$, however, could source unexpected effects
and, although equations have been set \cite{MM}, only their numerical
treatment can provide a reliable answer. According to the conjecture,
however, the fading of $\beta$--coupling could then trigger BH
production.

Let us finally comment about the stage B (before the onset of
radiative eras). SC cosmologies however set a bridge between the late
inflationary stages and our epoch. During this period, the $\Phi$
field underwent just a logarithmic growth, its present value being
just $\cal O$$(60)$ times its value at inflation end. Meanwhile,
kinetic energy has decreased by $\sim 240$ o.o.m.'s. The transition of
DE state parameter $w$ from +1 to -1, therefore, can be due to the
same potential causing the inflationary expansion only if it exhibits
an exponential dependence on $\Phi$.  This is somehow reminiscent of
the interaction $\cal L$$_I$ (eq.~1) which, anyhow, cannot be
responsible for the potential energy. The shape of $\cal L$$_I$
however tells us that a large field inflation might occur when $\Phi
\sim m_p \gg m$ (requiring then $b \gg 1$), so that the $\Phi$--$\psi$
interaction is practically switched off. A progressive decrease of
$\Phi$ could then reactivate $\cal L$$_I$ causing a reheating, when
$\Phi$ turns into $\psi$ quanta, to finally stabilize on the attractor
solution.  It might be appealing to consider this perspective also
within the frame of primeval particle production and annihilation due
to curvature variation \cite{zel}.

\vskip .2truecm
\noindent
{\bf Acknowledgments.} Ilia Musco is supported by postdoctoral funding
from ERC-StG EDECS no. 279954

\vfill\eject





\nocite{*} \bibliographystyle{elsarticle-num} \bibliography{martin}







\end{document}